\documentclass[11pt,twoside]{article}
\usepackage{./asp2014}

\aspSuppressVolSlug
\resetcounters

\begin{document}

\title{Magnetic White Dwarfs with Heavy Elements}
\author{François Hardy,$^1$ Patrick Dufour,$^1$ and Stefan Jordan$^2$
\affil{$^1$Université de Montréal, Montréal, QC, Canada; \email{hardy@astro.umontreal.ca}}
\affil{$^2$ ARI/ZAH, Universität Heidelberg, Heidelberg, Germany; \email{jordan@ari.uni-heidelberg.de}}}

\paperauthor{François Hardy}{hardy@astro.umontreal.ca}{}{Université de Montréal}{Département de physique}{Montréal}{QC}{H3T 1J4}{Canada}
\paperauthor{Patrick Dufour}{dufourpa@astro.umontreal.ca}{}{Université de Montréal}{Département de physique}{Montréal}{QC}{H3T 1J4}{Canada}
\paperauthor{Stefan Jordan}{jordan@ari.uni-heidelberg.de}{}{Universität Heidelberg}{Zentrum für Astronomie}{Heidelberg}{}{69120 }{Germany}

\begin{abstract}
Using our newly developed model atmosphere code appropriate for magnetic white dwarfs with metal lines in the Paschen-Back regime, we study various magnetic white dwarfs and explore the effects of various parameters such as the field geometry and the convective efficiency
\end{abstract}

\section{Effect of the Magnetic Field and Geometry}

For very high magnetic fields, the position and strength of the various splitted components can significantly differ to that calculated with simple Zeeman splitting. For illustrative purposes, we made an animation showing how some spectral lines evolve with increasing uniform magnetic fields, with one frame shown in Figure \ref{f:CaHK}. The geometry of the field also has a big impact on the line profile's shape, as seen in Figure \ref{f:inclination}. Full animations may be viewed at \mbox{\footnotesize \url{http://www.astro.umontreal.ca/~hardy/eurowd16/eurowd16.html}}.

In Figure \ref{f:inclination} we see the $\sigma$ components have very asymmetric profiles, something that isn't observed in magnetic DZ white dwarfs (Hollands et. al.). This asymmetry indicates that the surface magnetic fields appear to be more or less constant (similar to that in Figure \ref{f:CaHK}) or that the geometry is more complex than a simple dipole.

\articlefigure{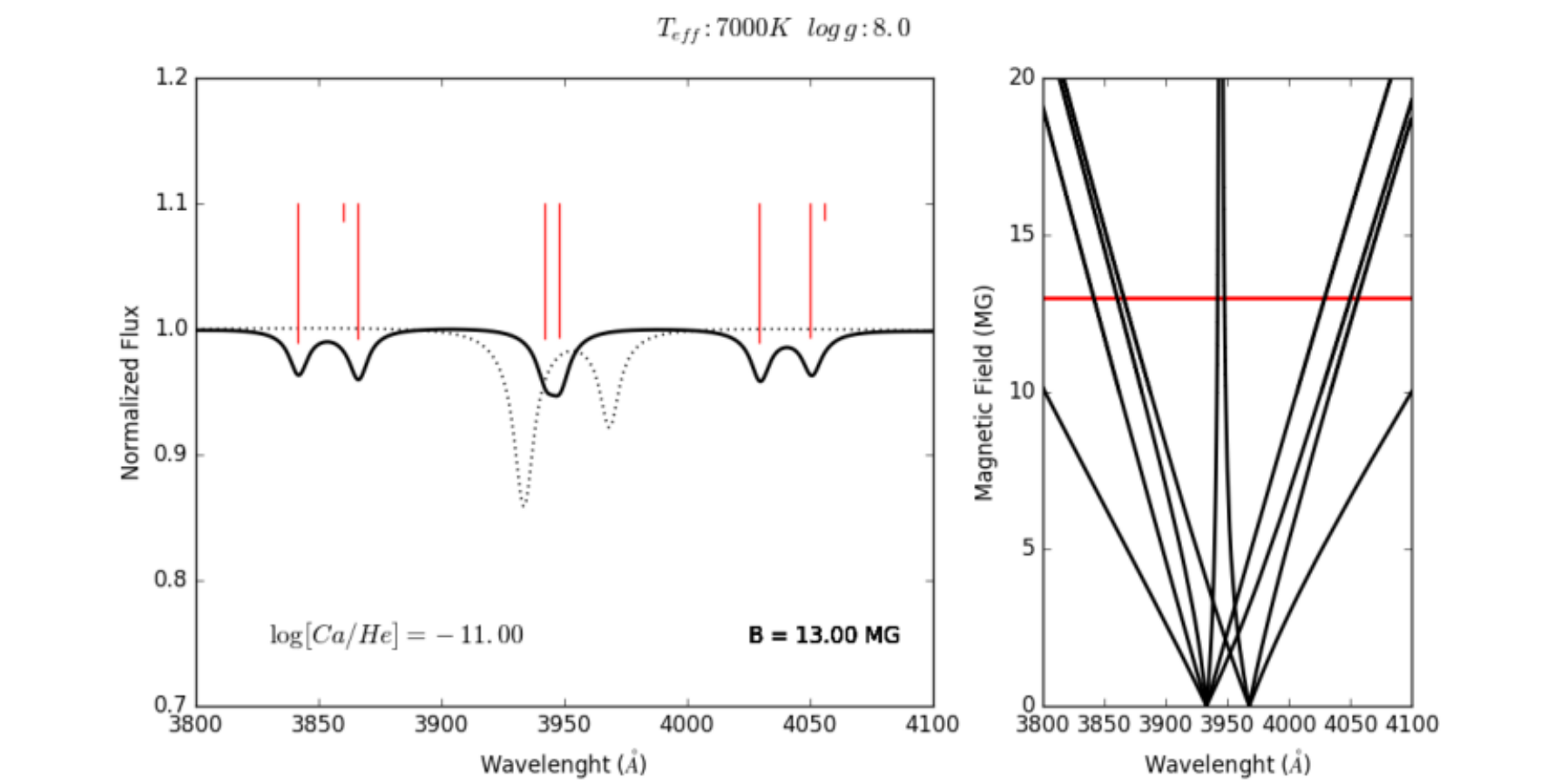}{f:CaHK}{One frame of our animation of the CaII H \& K line splitting in the Paschen-Back regime under an increasing and uniform magnetic field. Right panel: position of each line component as a function of magnetic field. Left panel: full line corresponds to the spectrum at the magnetic  field strength indicated by the red line in the right panel while the dotted line represent the B = 0 spectrum. Red ticks are proportional to the line strength of each component and vary as the magnetic field increases.}

\articlefigure{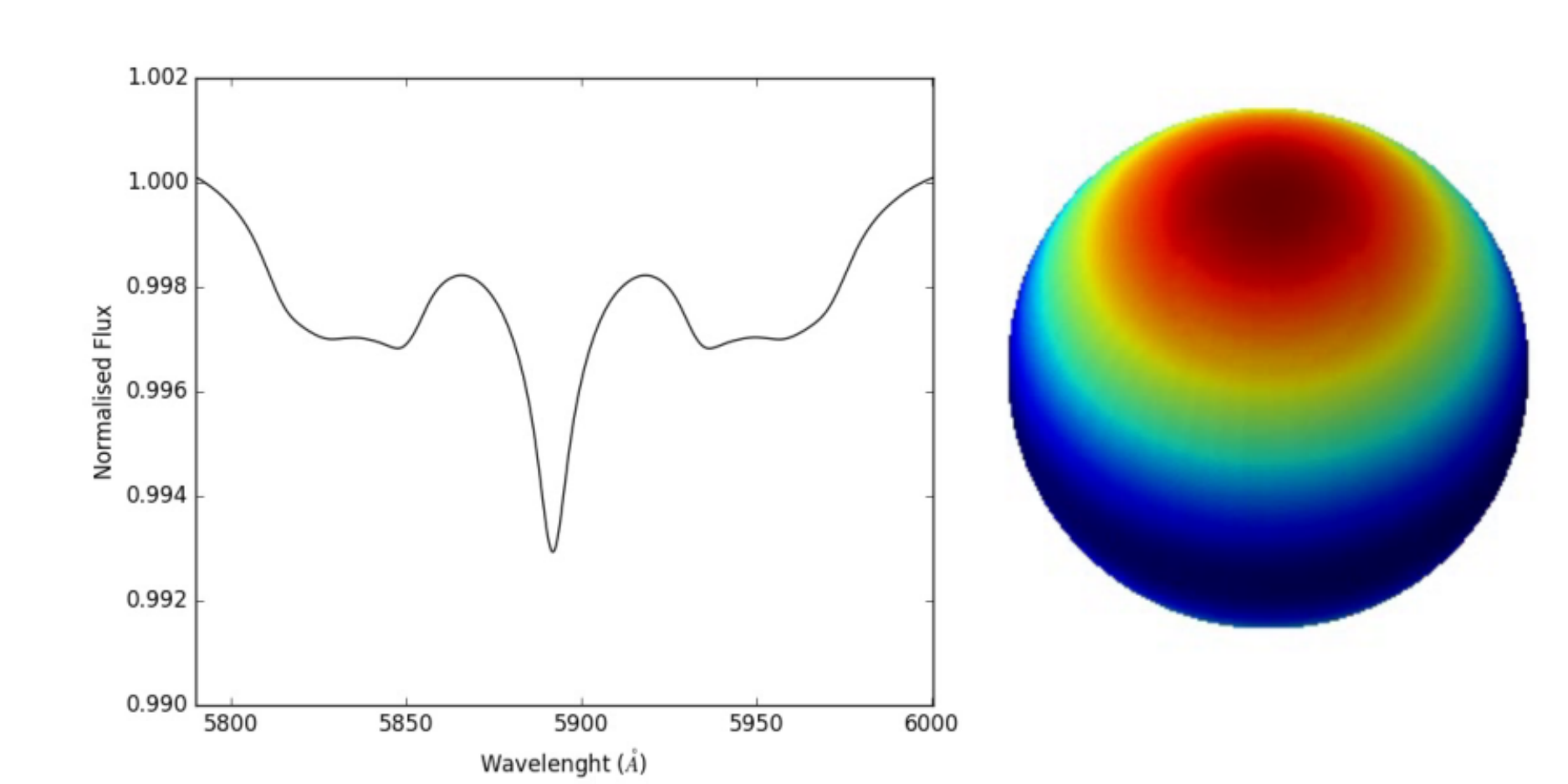}{f:inclination}{One frame of our animation showing the variation of the NaD doublet profile with the inclination of the magnetic dipole. The sigma components are broad and asymmetric, contrary to what is observed in known magnetic DZ white dwarfs (Holland et al. 2015). Only a large dipole offset (or a more or less constant surface field) can produce sharp sigma features, probably indicating that the true magnetic field structure is more complex than a simple dipole in cool DZ white dwarfs.}

\section{LHS 2534}

\articlefiguretwo{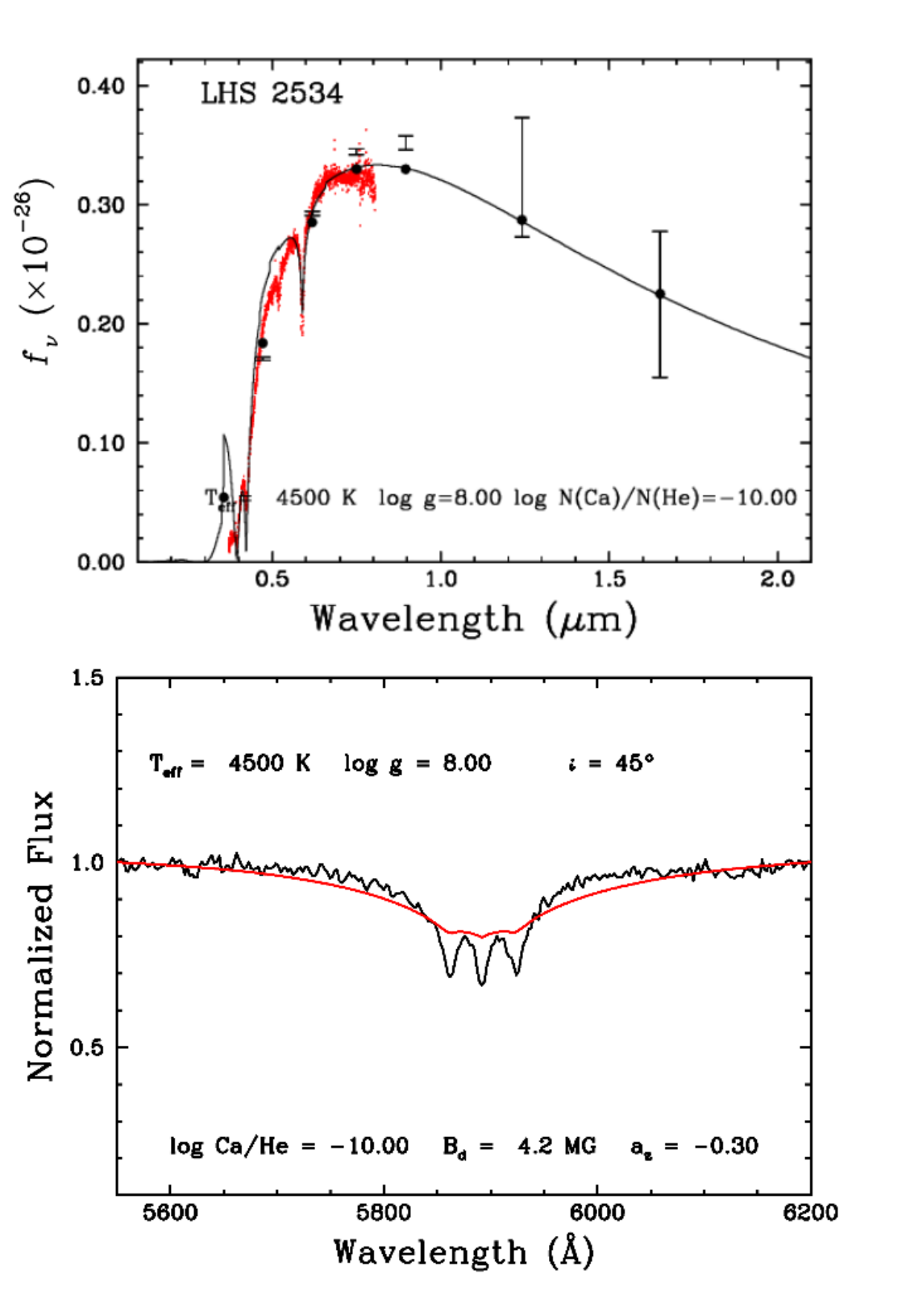}{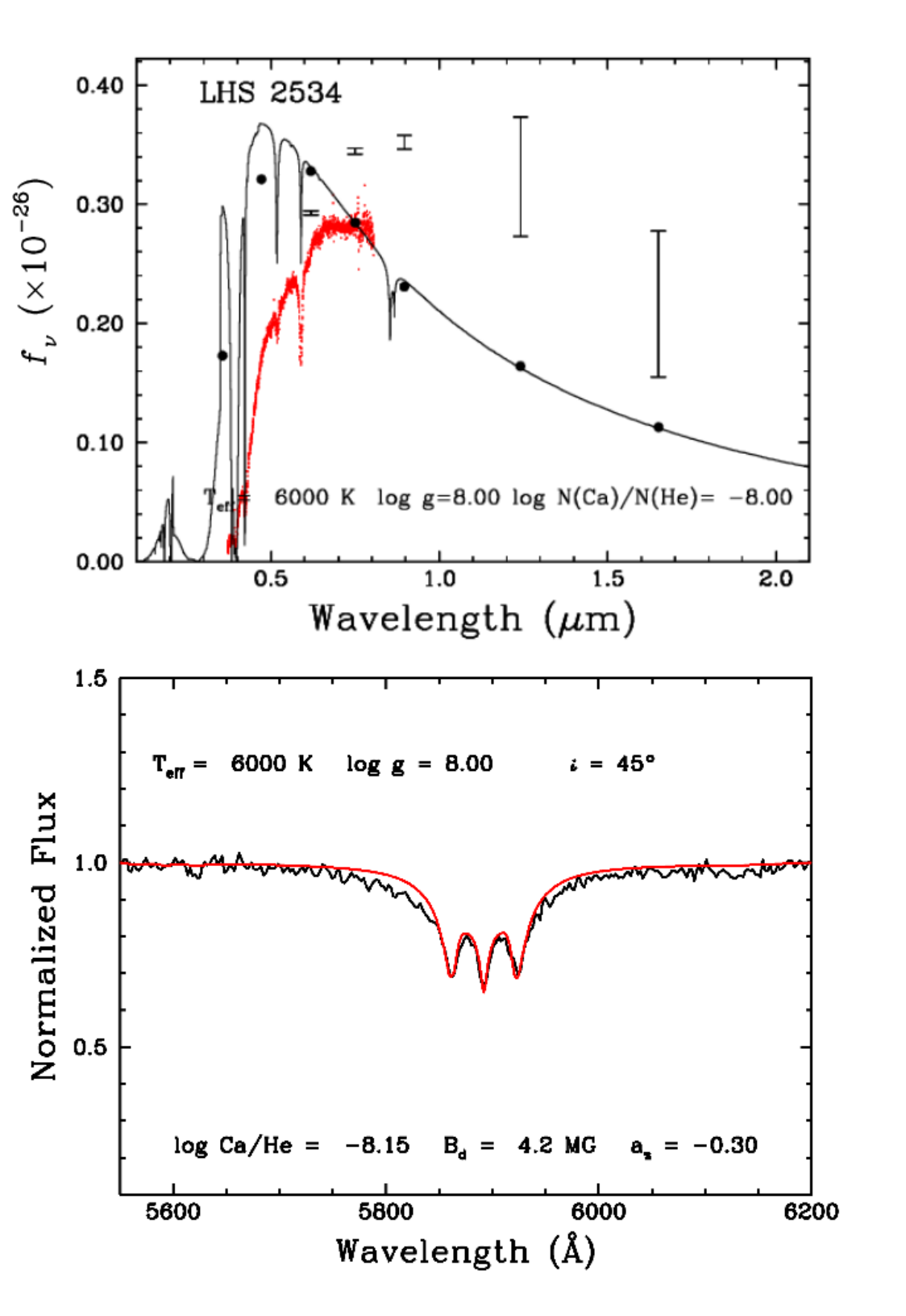}{f:photo_spec}{Top left panel: Fit to the ugriz+JH photometry with He+Z models. Bottom left panel: Best fit to the sodium NaD line assuming the photometric temperature fit. No combination of different $\log(g)$ and geometry produce better results. Right panel: Best fit to the NaD line using an offset magnetic dipole (by 0.3 stellar radii) and assuming $\log(g)$ = 8 and $T_{eff}$ = 6000 K. While the fit is much better than in the left panel, the corresponding photometric fit is not very good.}

Figure \ref{f:photo_spec} shows our photometric and spectroscopic analysis of LHS 2534\footnote{See \url{http://montrealwhitedwarfdatabase.org/WDs/LHS\%202534/LHS\%202534.html} and references therein.} . We have no explanation for the discrepancy between the photometric and spectroscopic fits. It is possible shortcomings in our models due to high pressure effects (see Simon Blouin's proceedings) or the high magnetic field (convection? see below) or non-uniform metal distribution on the surface (spots?) may be to blame.  A detailed exploration of these possibilities will be presented elsewhere.

\section{Convection}

\articlefigure[width=.75\textwidth]{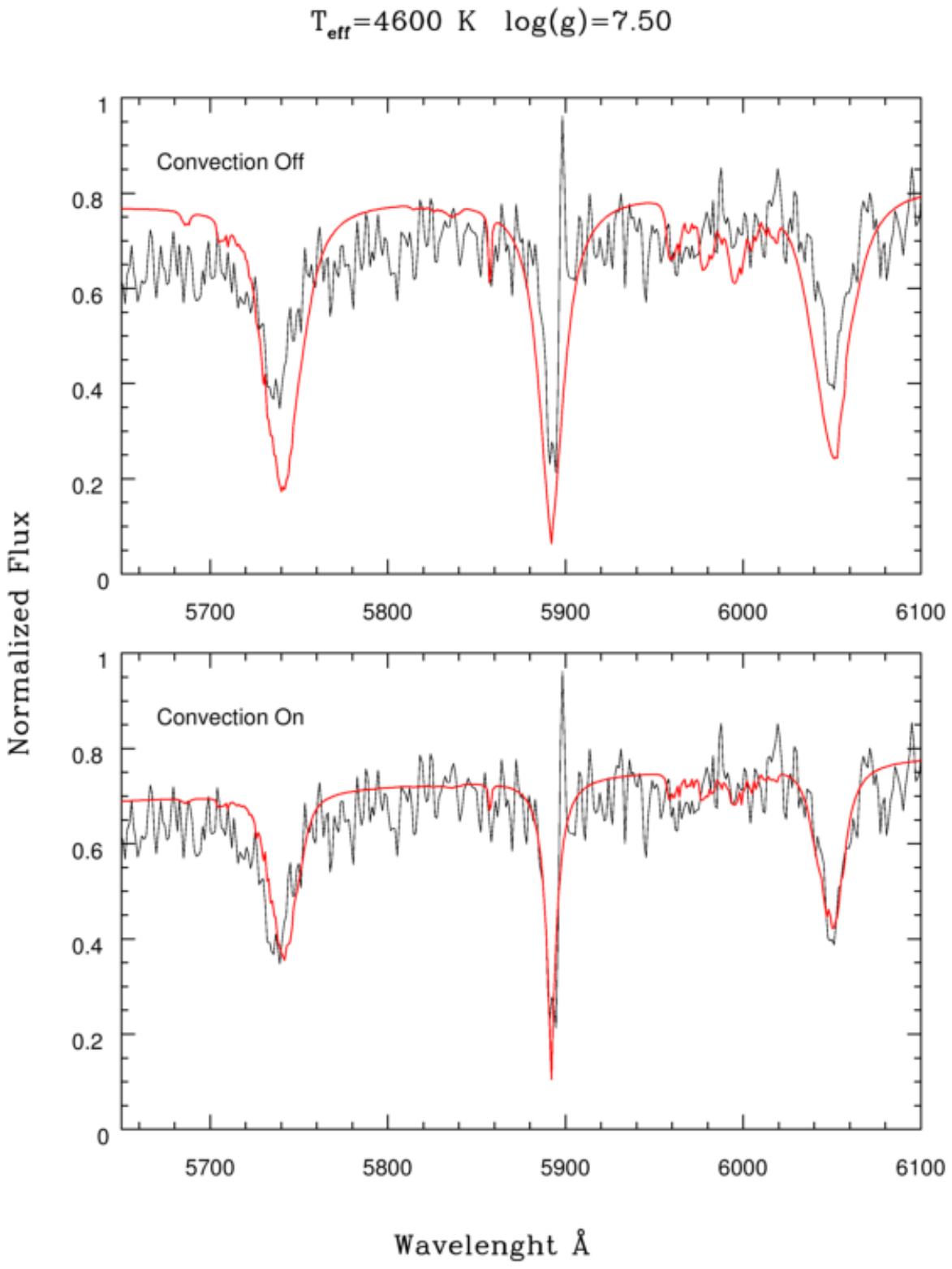}{f:spectres}{Best fit to the sodium lines of the magnetic DZ white dwarf J2105 (SDSS J2105+0900, Liebert et al. 2007). Upper panel: radiative models (convection off). Lower panel: standard models (convection on). We based our temperature and gravity parameters on previous results.}

Based on radiative magnetohydrodynamics simulations, Tremblay et al. 2015 recently suggested that convective energy transfer is seriously impeded in the presence of a magnetic field, with fields as low as 50 kG sufficient to completely stop convection. Assuming Dufour et. al. 2015 atmospheric parameters for the H-rich DZ SDSS J2105+0900, we explore the differences between a purely radiative atmosphere models, where convection is artificially removed, and standard convective models (note that no non-ideal effects due to high pressure are expected for this H dominated object).

Figure \ref{f:spectres} shows that spectral lines are much broader when convection is off (the pressure in the line forming region is much higher in a radiative model). Unless the surface gravities of all magnetic white dwarf are ridiculously low, this could indicate that even in the presence of high magnetic fields, convection is not completely inhibited and that the thermal structure would still be similar to that of a normal convective model.

\pagebreak
\acknowledgments We acknowledge support from NSERC (Canada) and FRQNT (Québec).


\begin{thebibliography}{}
\bibitem[Tremblay et al.(2015)]{2015ApJ...812...19T} Tremblay, P.-E., Fontaine, G., Freytag, B., et al.\ 2015, \apj, 812, 19 
\bibitem[Hollands et al.(2015)]{2015MNRAS.450..681H} Hollands, M.~A., G{\"a}nsicke, B.~T., \& Koester, D.\ 2015, \mnras, 450, 681 
\bibitem[Reid et al.(2001)]{2001ApJ...550L..61R} Reid, I.~N., Liebert, J., \& Schmidt, G.~D.\ 2001, \apjl, 550, L61 
\bibitem[Liebert et al.(2007)]{2007ASPC..372..129L} Liebert, J., Kilic, M., Williams, K.~A., et al.\ 2007, 15th European Workshop on White Dwarfs, 372, 129 
\bibitem[Dufour et al.(2015)]{2015ASPC..493...37D} Dufour, P., Jordan, S., Blouin, S., et al.\ 2015, 19th European Workshop on White Dwarfs, 493, 37 

\end{thebibliography}


\end{document}